\documentclass[journal, onecolumn]{IEEEtran}


\usepackage{cite}

\ifCLASSINFOpdf
   \usepackage[pdftex]{graphicx}
   \DeclareGraphicsExtensions{.pdf}
\else
   \usepackage[dvips]{graphicx}
   \DeclareGraphicsExtensions{.eps}
\fi

\usepackage[cmex10]{amsmath}

\usepackage{amssymb}

\DeclareMathOperator*{\maximize}{max.}

{}

\usepackage{suffix}
\usepackage{mathtools}
\DeclarePairedDelimiterX\MeijerM[3]{\lparen}{\rparen}%
{\begin{smallmatrix}#1 \\ #2\end{smallmatrix}\delimsize\vert\,#3}

\newcommand\MeijerG[8][]{%
  \mathsf{G}^{#2,#3}_{#4,#5}\MeijerM[#1]{#6}{#7}{#8}}

\WithSuffix\newcommand\MeijerG*[7]{%
  \mathsf{G}^{#1,#2}_{#3,#4}\MeijerM*{#5}{#6}{#7}}
	
\WithSuffix\newcommand\MeijerE*[7]{%
  \mathsf{E}\MeijerM*{#5}{#6}{#7}}

\usepackage[tight,footnotesize]{subfigure} 

\begin{document}
%
\title{Performance Analysis of Mean Value-Based Power Allocation with Primary User Interference in Spectrum Sharing Systems}

\author{Ruifeng~Duan,~\IEEEmembership{Member,~IEEE,}
        Riku~J\"{a}ntti,~\IEEEmembership{Senior Member,~IEEE,}
        and~ Mohammed~S.~Elmusrati,~\IEEEmembership{Senior Member,~IEEE}
\thanks{Manuscript on \today} 
}

\maketitle

\begin{abstract}
In this paper, we provide an exact expression for the ergodic capacity of the secondary user, and a unified closed-form expression for its bounds with taking the consideration of the primary interference at the secondary receiver. In addition, a simple but accurate approximation of the the ergodic capacity of the primary user is presented. Moreover, a primary user capacity loss based power allocation scheme for the secondary user is also proposed. Finally, we compare the performance of the two power allocation schemes in terms of the sum capacity. The results are validated through using numerical results and simulations. 

\end{abstract}

\begin{IEEEkeywords}
	Cognitive radio, mean-value power allocation, ergodic capacity.
\end{IEEEkeywords}

\IEEEpeerreviewmaketitle

\section{Introduction}

Cognitive radio (CR) systems operate much differently from the traditional wireless communication systems that the impact by transmission of secondary users (SUs or CR users) to a primary user must be controlled according to the primary user's requirements \cite{Goldsmith2009,Haykin2005}. The primary users may have strict constraints, for instance, the instantaneous interference at the a primary receiver generated by the SUs should be below a predefined threshold; some moderate constraints that only the average interference over all the channel states needs to satisfy a threshold, or the interference outage probability is controlled; the ergodic capacity loss of a primary user constraint where the capacity loss of the PU caused by the secondary transmission is predefined.

Power allocation acts as an effective way to control the secondary transmission in order to satisfy the primary users' requirements. There is a growing body of literature on power control/allocation in CR systems. In \cite{Chen2008c}, the secondary transmitter adjusts its transmission power to maximize its data rate without increasing the outage probability at the primary receiver. A soft-sensing based optimal power control scheme to maximize the secondary user (SU) capacity was proposed in \cite{Srinivasa2010}, where the peak interference power constraint was implemented to govern the secondary transmit power. \cite{Islam2008} and \cite{Qian2007} investigated the power control strategies for opportunistic spectrum access (OSA) in TV bands. For the interference control of the secondary users over television white spaces, Koufos \emph{et al.} in \cite{Koufos2011} proposed the power density and deployment based transmit power control of the secondary users such that the quality of the TV services is not violated by the aggregated interference.

With perfect instantaneous channel information, many works have studied the optimal transmit power allocation schemes in order to maximize the capacity of the secondary system under long-term and/or short-term constraints. The seminal work in \cite{Gastpar2004, Gastpar2007} first investigated the ergodic capacity of different non-fading additive-white-Gaussian noise (AWGN) channels, where the transmit power of the SU was regulated by the average interference power received at a third-party receiver. The significant finding is that the received-signal constraints can lead to substantially different results as compared to transmitted-signal constraints. This result has inspired the research work on optimal power allocation for SUs over fading channels. Results in \cite{Ghasemi2007} showed that over fading channels significant capacity gains may be achieved by the SUs. In \cite{Suraweera2008}, the authors extended the work in \cite{Ghasemi2007} by investigating the achievable capacity gains in asymmetric fading environments.

Thereafter, the SU capacity gains by spectrum-sharing approach in a Rayleigh fading environment were investigated in \cite{Musavian2009}. The optimal power allocation strategies for a SU achieving the ergodic, delay-limited, or outage capacity were proposed in \cite{Kang2009}. The authors observed that fading of the channel from secondary transmitter to primary receiver can be a good phenomenon for maximizing the capacity of the SU. Zhang concluded in \cite{Zhang2009d} that the average-interference-power (AIP) constraint can be more advantageous over the peak-interference-power for minimizing the resultant capacity loss of the primary fading channel. Having a further study, \cite{Zhang2008a} showed that the primary user capacity loss constraint is superior to other constraints in general, where all channels' information were available. However, in the previous mentioned works, the interference from the primary user to the secondary user or the noise at the secondary receiver was neglected due to difficulties to obtain the final expressions.

Different from the previous power allocation strategies based on the perfect instantaneous channel state information, several works, for instance, \cite{Suraweera2010, Rezki2012, Lim2012,Smith2013}, investigated the performance of spectrum-sharing systems with limited channel knowledge. \cite{Suraweera2010} studied the SU mean capacity of a spectrum-sharing system under the peak interference power constraint predefined by the PU and the secondary peak transmit power constraint. The authors considered the impact of the interference from the PU transmission. The analytical expressions for the cumulative distribution functions of the SU capacity were investigated in \cite{Smith2013}. However, the analytical results were provided consisting of integrals. \cite{Rezki2012, Lim2012} have studied the mean-value based power allocation for secondary users, where the SU has only the statistical information of the channel ST-PR. Nevertheless, the impact of the interference from the PU was not taken into consideration.

In this paper, we study the performance of the secondary user in terms of ergodic capacity by taking into consideration of two constraints predefined by the primary user or regulators, interference outage probability at the primary user and primary ergodic capacity loss percentage. The interference of the primary user to the secondary transmission is considered, which has not been, to the best of our knowledge, taking into consideration in the literature. We found that the interference of the primary user to the secondary transmission has different influences on the secondary transmit power for the two scenarios studied in this paper.

We highlight the main contributions of this paper in the following:
\begin{itemize}
	\item An exact expression for the secondary capacity with the consideration of the interference from the primary user.
	\item Closed-form expressions of the tight bounds for the secondary ergodic capacity.
	\item An accurate approximation for the ergodic capacity of the primary user.
\end{itemize}

The remainder of this paper is organized as follows. In Section \ref{sec:model}, we illustrate the system and channel models. In Section \ref{sec:mv_pa_cap}, we study the mean value-based power allocation strategy and the related ergodic capacity for the secondary user. We first derive the exact expression for the secondary ergodic capacity but with infinite summation, and then obtain a unified closed-form expression for the upper and lower bound of the ergodic capacity. Section \ref{sec:cap_los_pu} analyzes the primary user ergodic capacity loss due to secondary transmission using the power allocation scheme in Sec. \ref{sec:mv_pa_cap}. We provide a unified closed-form expression for the bounds of the PU capacity loss, and then obtain an accurate but simple closed form expression. Inspired from the results in previous sections, we propose in Section \ref{sec:pu_c_los_pa} a PU capacity loss-based power allocation scheme for the secondary user. Analytical and simulated results are then presented in each above section. The last section concludes this paper.

\emph{Notations:}  The expectation operation is denoted by $\mathbb{E}\left\{\cdot\right\}$. $\left[x\right]^+$ denotes $\max(0,x)$. $\overline{x}$ represents the mean value of $x$. $\mathsf{E_i}(x) = -\int_{-x}^{\infty}\frac{e^{-t}}{t}\, dt, \forall x<0$ denotes the exponential integral function \cite[eqn. 8.211]{Gradshteyn2007}, and $\Gamma(\alpha,x)=\int_x^{\infty}e^{-t}t^{\alpha-1}dt$ is the incomplete gamma function \cite[eqn. 8.350]{Gradshteyn2007}. $\log$ denotes natural logarithm operation in this paper.

\section{System Model} \label{sec:model}

\begin{figure}[!t]
	\centering
		\includegraphics[width=0.80\columnwidth]{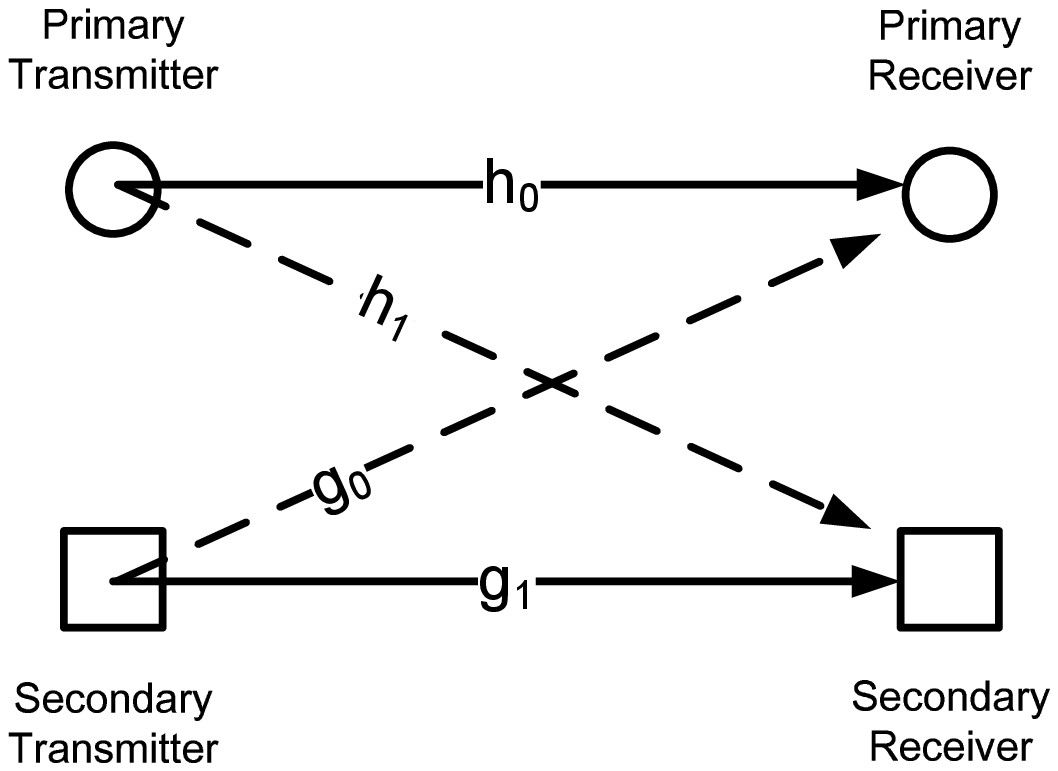}
	\caption{System model.}
	\label{fig:SysMod}
\end{figure}

We consider a spectrum sharing scenario in paper which is depicted in Fig. \ref{fig:SysMod}. The primary user (PU) is working on a licensed frequency band. The secondary user (SU) coexists with the PU sharing the same spectrum under the predefined constraints by the PU. The secondary transmitter (ST) has no the instantaneous channel state information of the ST-PR link but the statistical mean value. In addition, we assume that the PU transmits using fixed transmit power. We consider independent additive white Gaussian noise (AWGN) block-fading channels. The block-fading, or quasi-static, channel model was introduced in \cite{Ozarow1994} and has been commonly using for studying wireless communications systems over slowly-varying fading channels \cite{Ozarow1994, Biglieri1998}. During each fading block, the channel gain remains constant while varying from block to block. Moreover, all channels experience Rayleigh fading, such that the channel power gains are exponentially distributed. For an exponential variable $Z$, the probability density function is given by 
\begin{equation}
	f_{Z}(z)=\frac{1}{\overline{z}}e^{-\frac{z}{\overline{z}}}
\end{equation}
where $\overline{z}$ denotes the mean value of $Z$.

\section{Mean Value-based Power Allocation and Ergodic Capacity} \label{sec:mv_pa_cap}

The ergodic capacity under mean-value based power allocation (MVPA) can be achieved through employing a frame work presented in \cite{Rezki2012}. The ergodic capacity optimization problem based on MVPA for the SU can be formulated as \cite{Lim2012}
\begin{IEEEeqnarray}{rl}
	C &= \maximize_{P_s(\overline{g}_0)} \quad \mathbb{E}_{g_1,h_1}\left[\log\left(1+\frac{g_1P_s(\overline{g}_0)}{P_p h_1 + N_S}\right)\right]  \\
 \text{s.t.} & \quad 1-\Pr\left\{g_0 P_s(\overline{g}_0)\leq I_{th}\right\}\leq P_O^{th}
\end{IEEEeqnarray}
where $P_s(\overline{g}_0)$ denotes the transmit power of the SU, $P_p$ represents the transmit power of the PU, $I_{th}$ is the predefined interference threshold by the PU, $P_O^{th}$ is the predefined, by the PU, interference outage probability that the instantaneous interference exceeds $I_{th}$, and $N_S$ is the additive white Gaussian noise power at the SR.

According to \cite{Rezki2012}, the interference outage probability constraint in the above optimization problem is equivalent to 
\begin{equation}
	P_s(\overline{g}_0) \leq \frac{I_{th}}{F_{g_0}^{-1}(1-P_O^{th})} \label{P_s_IO}
\end{equation}
where $F_{g_0}^{-1}(1-P_O^{th})$ is the inverse cumulative distribution function of $g_0$, and for Rayleigh fading scenarios, the probability density function of the channel power gain is continuous and not null so that $F_{g_0}^{-1}(\cdot)$ exists.

In the MVPA, the secondary transmitter has no instantaneous ST-PR channel state information, so that the ST relies on the statistics of ST-PR and its own channel information, $g_1$, \cite{Lim2012}. In addition, $F_{g_0}^{-1}(1-P_O^{th})$ takes a fixed value \cite{Rezki2012}. This means that the secondary user uses fixed transmit power to achieve the ergodic capacity under the interference outage probability constraint, such that (\ref{P_s_IO}) gives the maximum transmit power of the SU. Based on the setting that $g_0$ is exponentially distributed with a mean of $\overline{g}_0$, i.e., $F_{g_0}^{-1}(1-P_O^{th}) = \overline{g}_0 \log\left(\frac{1}{P_O^{th}}\right)$. Consequently, the fixed transmit power for the secondary transmitter is 
\begin{equation}
	P_s(\overline{g}_0) = \frac{I_{th}}{\overline{g}_0 \log\left(\frac{1}{P_O^{th}}\right)}
\end{equation}
In the following we omit the parameter $\overline{g}_0$, i.e., $P_s$ is used. We can obtain the ergodic capacity of the SU under MVPA as follows
\begin{IEEEeqnarray}{rl}
	C & = \int_0^{\infty} \int_0^{\infty} \log\left(1+\frac{g_1P_s}{P_ph_1 + N_S}\right)f_{g_1}(g_1) f_{h_1}(h_1)\,dg_1 dh_1  \nonumber\\
	  & = \int_0^{\infty}\int_0^{\infty} \log\left(1+\frac{g_1P_s}{P_ph_1 + N_S}\right)\frac{e^{-\frac{g_1}{\overline{g}_1}}}{\overline{g}_1} \frac{e^{-\frac{h_1}{\overline{h}_1}}}{\overline{h}_1} \,dg_1 dh_1   \nonumber\\
		& = \int_0^{\infty}  e^{\frac{P_P h_1 + N_S}{\overline{g}_1P_s}}\Gamma\left(0,\frac{P_P h_1 + N_S}{\overline{g}_1P_s}\right)  \frac{1}{\overline{h}_1}e^{-\frac{h_1}{\overline{h}_1}} \, dh_1  \label{erg_cap_1}
\end{IEEEeqnarray}
where in the last two steps we have used \cite[4.337-2 and 8.359-1]{Gradshteyn2007}. To the best of our knowledge, there is no direct way to find the final expression for the integral in (\ref{erg_cap_1}). Therefore, we present, in the following, a result for (\ref{erg_cap_1}) in terms of infinite summation, and then a closed-form expression for the unified upper and lower bounds of the the secondary capacity is proposed.

First, we derive an expression with infinite summation using the following relation
\begin{equation}
	e^{x}\Gamma(0,x) = \sum_{k=0}^{\infty}\frac{L_k(x)}{k+1}
\end{equation}
where $L_k(x) = \sum_{m=0}^k(-1)^{k+m}\begin{pmatrix}k\\k-m\end{pmatrix}\frac{x^m}{m!}$
is the Laguerre polynomial expression given by \cite[eqn. 8.354-5, 8.970-1 and 8.970-2]{Gradshteyn2007}. Therefore, (\ref{erg_cap_1}) can be written as
\begin{IEEEeqnarray}{rl}
	C & = \sum_{k=0}^{\infty} \frac{1}{k+1} \sum_{m=0}^k(-1)^{k+m}\begin{pmatrix}k\\k-m\end{pmatrix} \nonumber\\
			& \quad \times\int_0^{\infty} \frac{\left(\frac{P_P h_1 + N_S}{\overline{g}_1P_s}\right)^m}{m!}  \frac{1}{\overline{h}_1}e^{-\frac{h_1}{\overline{h}_1}} \, dh_1 \\
		& = \sum_{k=0}^{\infty} \frac{1}{k+1} \sum_{m=0}^k(-1)^{k+m}\begin{pmatrix}k\\k-m\end{pmatrix}  \frac{1}{m!}\left(\frac{P_P \overline{h}_1}{\overline{g}_1P_s}\right)^m \nonumber\\
			& \quad \times e^{\frac{N_S}{P_P \overline{h}_1}} \Gamma\left(m+1, \frac{N_S}{P_P \overline{h}_1}\right) \label{Cap_sum_1}
\end{IEEEeqnarray}
where we used \cite[eqn. 3.351-2]{Gradshteyn2007} in the last step. However, this expression does not show some insights explicitly.

\begin{figure}[htbp]
	\centering
		\includegraphics[width=0.80\columnwidth]{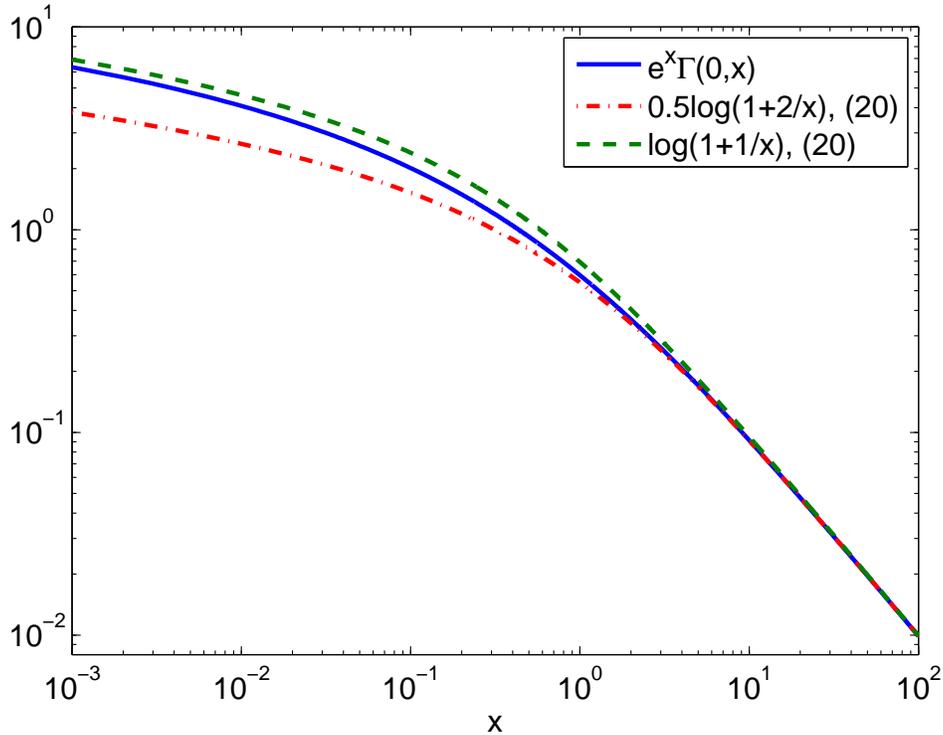}
	\caption{Illustration of $e^x\Gamma(0,x)$ and its bounds.}
	\label{fig:bounds_x}
\end{figure}

Then, we derive a unified closed-form expression for the bounds of the secondary ergodic capacity in (\ref{erg_cap_1}). From the expression in (\ref{erg_cap_1}), we can see that the integration over the interference channel PT-SR consists of the term $e^x\Gamma(0,x)$. To the best of our knowledge, there is no closed-form expression for the ergodic capacity. Therefore, we consider the bounds of the incomplete Gamma function to obtain some expressions for (\ref{erg_cap_1}). The bounds of the product of an exponential and an incomplete Gamma function are given in \cite{Alzer1997} as 
\begin{equation}
	0.5 \log\left(1+\frac{2}{x}\right) \leq e^x\Gamma(0,x) \leq \log\left(1+\frac{1}{x}\right)
\end{equation}
We can have an unified form for the upper and lower bounds of $e^x\Gamma(0,x)$ as
\begin{equation}
	c \log\left(1+\frac{1}{cx}\right)
\end{equation}
where $c=0.5$ gives the lower bound, and $c=1$ for upper bound, which are illustrated in Fig.\ref{fig:bounds_x}.

The closed-form expression for the bounds of the secondary ergodic capacity becomes

\begin{IEEEeqnarray}{rl}
	& C^{(b)} \nonumber\\
		& = \int_0^{\infty}  e^{\frac{P_P h_1 + N_S}{\overline{g}_1P_s}}\Gamma\left(0,\frac{P_P h_1 + N_S}{\overline{g}_1P_s}\right)  \frac{1}{\overline{h}_1}e^{-\frac{h_1}{\overline{h}_1}} \, dh_1 \nonumber\\
		& = \int_0^{\infty}  c \log\left(1 + \frac{\overline{g}_1P_s/c}{P_P h_1 + N_S}\right) \frac{1}{\overline{h}_1}e^{-\frac{h_1}{\overline{h}_1}} \, dh_1 \nonumber\\
		& = c\int_0^{\infty} \left[\log\left(1+\frac{\overline{g}_1P_s}{c N_S}\right) + \log\left(1+\frac{P_P}{\overline{g}_1P_s/c+ N_S} h_1 \right) \right.\nonumber\\
			& \quad \left. - \log\left(1+\frac{P_P}{N_S}h_1\right)\right] \frac{1}{\overline{h}_1}e^{-\frac{h_1}{\overline{h}_1}} \, dh_1  \nonumber\\
		& = c\log\left(1+\frac{\overline{g}_1P_s}{c N_S}\right) - c\left[e^{\frac{N_S}{P_P \overline{h}_1}}\Gamma\left(0,\frac{N_S}{P_P \overline{h}_1}\right) \right. \nonumber\\
			& \quad \left. - e^{\frac{\overline{g}_1P_s/c + N_S}{P_P \overline{h}_1}}\Gamma\left(0,\frac{\overline{g}_1P_s/c + N_S}{P_P \overline{h}_1}\right)\right]
\end{IEEEeqnarray}
where in the last two steps we have used \cite[4.337-2 and 8.359-1]{Gradshteyn2007}. The last two terms in the above result illustrate the average capacity loss caused by the interference from the primary user.

\begin{figure}[!ht]
	\centering
		\includegraphics[width=0.90\columnwidth]{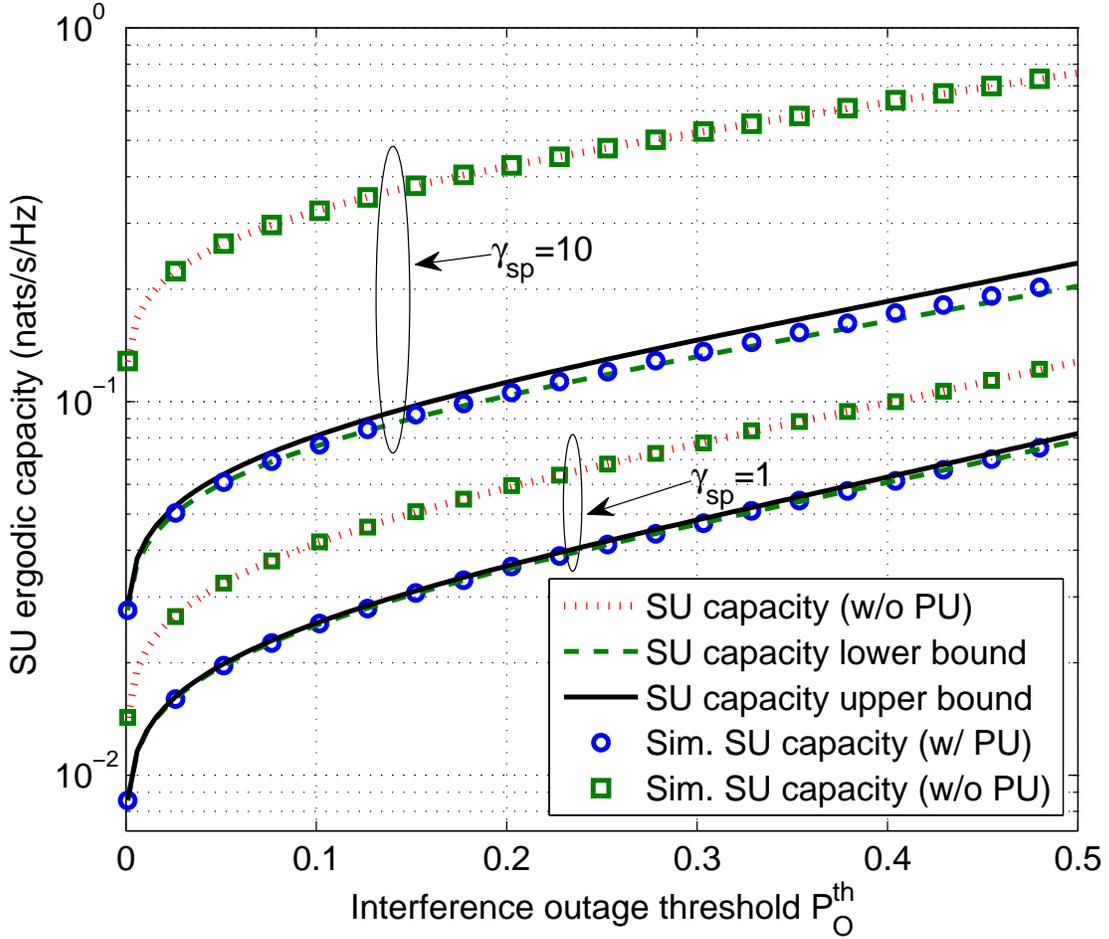}
	\caption{Secondary user ergodic capacity. $\gamma_{sp}=P_P\overline{h}_1/N_s$, $\overline{g}_1 = 1$, and $N_s = 1$. Interference threshold $I_{th}=0.1 \gamma_p$.}
	\label{fig:c_su}
\end{figure}

In addition, when the interference from the primary transmission is ignorable, we have the following limit
\begin{equation}
	\lim_{P_P \overline{h}_1 \rightarrow 0} C^{(b)} \rightarrow c\log\left(1+\frac{\overline{g}_1P_s}{c N_S}\right)
\end{equation}
where we used $\lim_{x \rightarrow \infty}\Gamma(a,x)\rightarrow 0$. This result verifies that when the interference from the PU is decreasing, the capacity loss of the SU is vanishing. It is also consistent with \cite[eqn. (11)]{Lim2012} as $c\log\left(1+\frac{\overline{g}_1P_s}{c N_S}\right)$ represents the bounds of $e^{\frac{N_S}{\overline{g}_1P_s}}\Gamma\left(0, \frac{N_S}{\overline{g}_1P_s}\right)$. Fig. \ref{fig:c_su} depicts the ergodic capacity of the SU, where we use the same transmission power for the SU in both two cases, with and without primary user. We define the interference threshold $I_{th}$ as 10\% of the SNR of the primary user, i.e., $I_{th}=0.1 \gamma_p = \frac{0.1 P_P\overline{h}_0}{N_p}$.

\section{Capacity loss of the primary user} \label{sec:cap_los_pu}
In the previous section, the impact of the secondary transmission on the primary user is characterized by the interference outage probability. Also, we can can see this is a strong constraint on the secondary transmission. Therefore, we may be interested in knowing, from another aspect, that how much affect in terms of the primary ergodic capacity loss the secondary transmission may cause. In this section, we derive the capacity loss of the PU due to secondary transmission. Without transmission of the SU, the capacity of the PU with fixed power transmission can be obtained as following,
\begin{IEEEeqnarray}{rl}
	C_P & = \int_0^{\infty}  \log\left(1 + \frac{h_0P_P}{N_P}\right) \frac{1}{\overline{h}_0}e^{-\frac{h_0}{\overline{h}_0}} \, dh_0 \nonumber\\
		  & = - e^{\frac{N_P}{P_P \overline{h}_0}} \text{Ei}\left(-\frac{N_P}{P_P \overline{h}_0}\right) 
\end{IEEEeqnarray}
where we follow the assumption of Rayleigh fading channels. Equivalently, $C_P$ can be written as
\begin{equation}
	C_P =  e^{\frac{N_P}{P_P \overline{h}_0}} \text{E}_1\left(\frac{N_P}{P_P \overline{h}_0}\right)  = e^{\frac{N_P}{P_P \overline{h}_0}}\Gamma\left(0,\frac{N_P}{P_P \overline{h}_0}\right)
\end{equation}

The ergodic capacity of the primary user with interference from the SU can be derived as
\begin{IEEEeqnarray}{rl}
	& C_P^{s} \nonumber\\
	& = \int_0^{\infty} \int_0^{\infty} \log\left(1 + \frac{h_0P_P}{P_s g_0 + N_P}\right) \frac{1}{\overline{h}_0}e^{-\frac{h_0}{\overline{h}_0}} \frac{1}{\overline{g}_0}e^{-\frac{g_0}{\overline{g}_0}}\, dh_0 dg_0 \nonumber\\
			& = \int_0^{\infty}  e^{\frac{P_s g_0 + N_P}{\overline{h}_0P_P}}\Gamma\left(0,\frac{P_s g_0 + N_P}{\overline{h}_0P_P}\right)  \frac{1}{\overline{g}_0}e^{-\frac{g_0}{\overline{g}_0}} \, dg_0 \label{C_p}
\end{IEEEeqnarray}
We can see that it is similar to (\ref{erg_cap_1}), and the integration can be done in the same way as in (\ref{Cap_sum_1}). Therefore, we omit the derivations, and derive its bounds in order to obtain some insights explicitly.

\begin{IEEEeqnarray}{rl}
	& C_P^{s,(b)} \nonumber\\
		& = \int_0^{\infty}  c\log\left(1+\frac{\overline{h}_0P_P/c}{P_s g_0 + N_P}\right)  \frac{1}{\overline{g}_0}e^{-\frac{g_0}{\overline{g}_0}} \, dg_0  \nonumber\\
		& = c\int_0^{\infty} \left[\log\left(1+\frac{\overline{h}_0P_P/c}{N_P}\right) + \log\left(1+\frac{P_s g_0}{\overline{h}_0P_P/c+ N_P}  \right) \right.\nonumber\\
			& \quad \left. - \log\left(1+\frac{P_s}{N_P}g_0\right)\right] \frac{1}{\overline{g}_0}e^{-\frac{g_0}{\overline{g}_0}} \, dg_0  \nonumber\\
		& = c\log\left(1+\frac{\overline{h}_0P_P/c}{N_P}\right) - c\left[e^{\frac{N_P}{P_s \overline{g}_0}}\Gamma\left(0,\frac{N_P}{P_s \overline{g}_0}\right) \right. \nonumber\\
			& \quad \left. - e^{\frac{\overline{h}_0P_P/c + N_P}{P_s \overline{g}_0}}\Gamma\left(0,\frac{\overline{h}_0P_P/c + N_P}{P_s \overline{g}_0}\right)\right] \label{C_p_bounds}
\end{IEEEeqnarray}
The same as in the previous section that the last two terms can be considered as the average capacity loss of the primary user due to the secondary transmission. The first term represents the maximum achievable rate without secondary transmission through using the relation in (\ref{fig:bounds_x}). Then we obtain the unified expression for the bounds of the capacity loss of the PU.

\begin{IEEEeqnarray}{rl}
	C_P^{\text{loss}} & = C_P-C_P^{s,(b)} \nonumber\\
		& = e^{\frac{N_P}{P_P \overline{h}_0}}\Gamma\left(0,\frac{N_P}{P_P \overline{h}_0}\right) - c\log\left(1+\frac{\overline{h}_0P_P/c}{N_P}\right)  \nonumber \\
			& \quad + c\left[e^{\frac{N_P}{P_s \overline{g}_0}}\Gamma\left(0,\frac{N_P}{P_s \overline{g}_0}\right) \right. \nonumber\\
			& \quad \left. - e^{\frac{\overline{h}_0P_P/c + N_P}{P_s \overline{g}_0}}\Gamma\left(0,\frac{\overline{h}_0P_P/c + N_P}{P_s \overline{g}_0}\right)\right]
\end{IEEEeqnarray}




\begin{figure}[!ht]
	\centering
		\includegraphics[width=0.90\columnwidth]{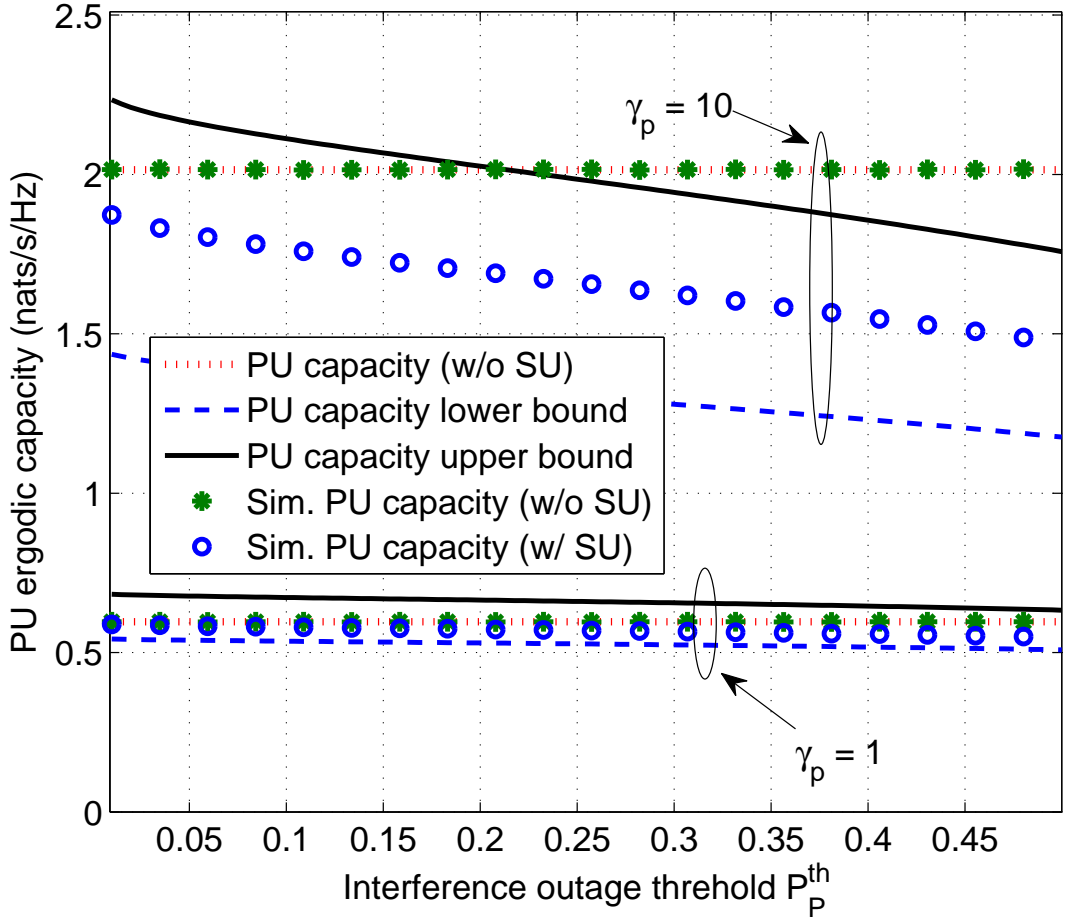}
	\caption{Primary user capacity. $\gamma_p = P_P\overline{h}_0/N_p$, $I_{th} = 0.1\gamma_p$, and $\overline{g}_0 = 1$.}
	\label{fig:c_pu_su}
\end{figure}

PU capacity versus interference outage probability is illustrated in Fig. \ref{fig:c_pu_su}.  However, because of the fact that the bounds of $e^x\Gamma(0,x)$ are quite loose when $x$ is a small value, for instance, the primary user is working in high SINR regime, the bounds of the primary user capacity are pretty far away from each other. From the observation of the bounds of $e^x\Gamma(0,x)$, we simply, by taking the arithmetic mean of the bounds, propose a approximation for it as following
\begin{equation}
	e^x\Gamma(0,x) \approx 0.5\left[\log\left(1+\frac{1}{x}\right) + 0.5 \log\left(1+\frac{2}{x}\right)\right]
\end{equation}

Then, we have the approximated PU ergodic capacity of (\ref{C_p}) could be written using (\ref{C_p_bounds}) as,
\begin{IEEEeqnarray}{rl}
	& C_P^{s,(\text{approx})} \nonumber\\
		& = 0.5\left[\log\left(1+\frac{\overline{h}_0P_P}{N_P}\right) - 1.5e^{\frac{N_P}{P_s \overline{g}_0}}\Gamma\left(0,\frac{N_P}{P_s \overline{g}_0}\right) \right. \nonumber\\
			& \quad + e^{\frac{\overline{h}_0P_P + N_P}{P_s \overline{g}_0}}\Gamma\left(0,\frac{\overline{h}_0P_P + N_P}{P_s \overline{g}_0}\right) + 0.5\log\left(1+\frac{2\overline{h}_0P_P}{N_P}\right) \nonumber \\
			& \quad \left. + 0.5e^{\frac{2\overline{h}_0P_P + N_P}{P_s \overline{g}_0}}\Gamma\left(0,\frac{2\overline{h}_0P_P + N_P}{P_s \overline{g}_0}\right)\right]\label{C_p_approx}
\end{IEEEeqnarray}

\begin{figure}[htbp]
	\centering
		\includegraphics[width=0.90\columnwidth]{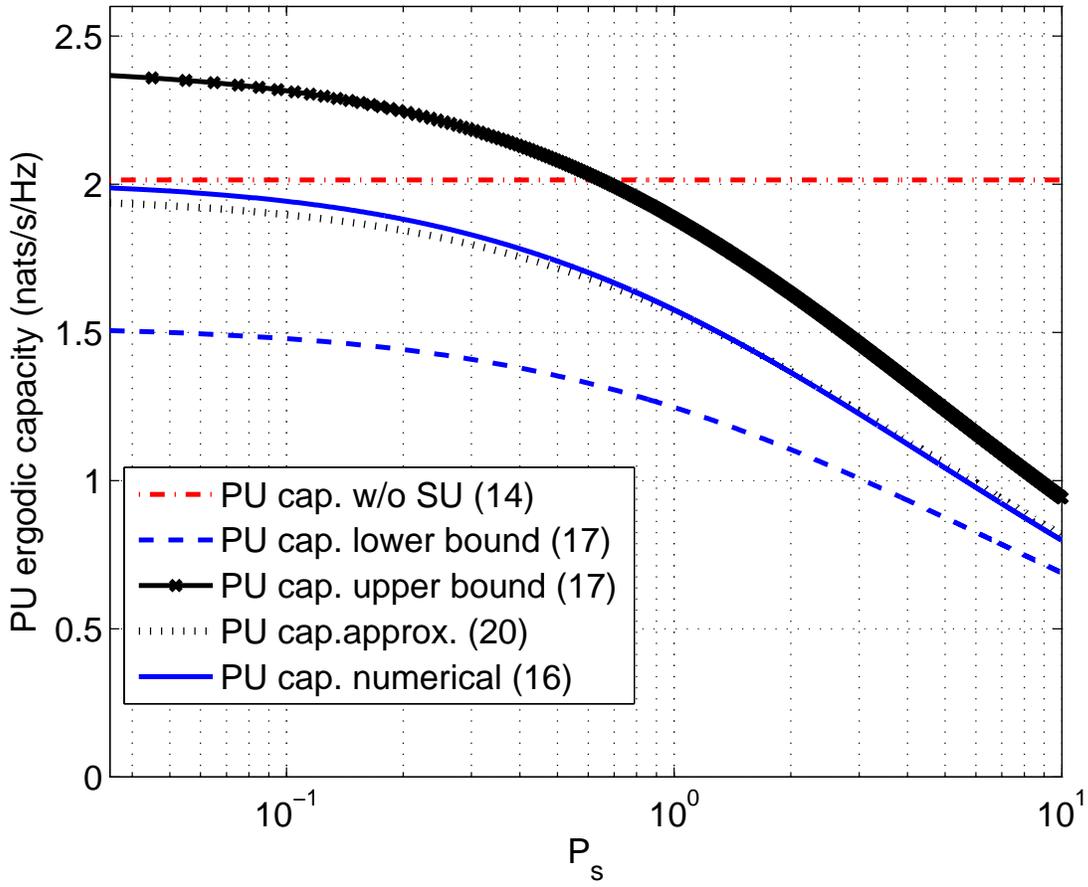}
	\caption{PU capacity comparison. $\gamma_p = 10$, $N_s=1$, $\overline{g}_0=1$.}
	\label{fig:c_pu_apprx}
\end{figure}

\begin{figure}[htbp]
	\centering
		\includegraphics[width=0.90\columnwidth]{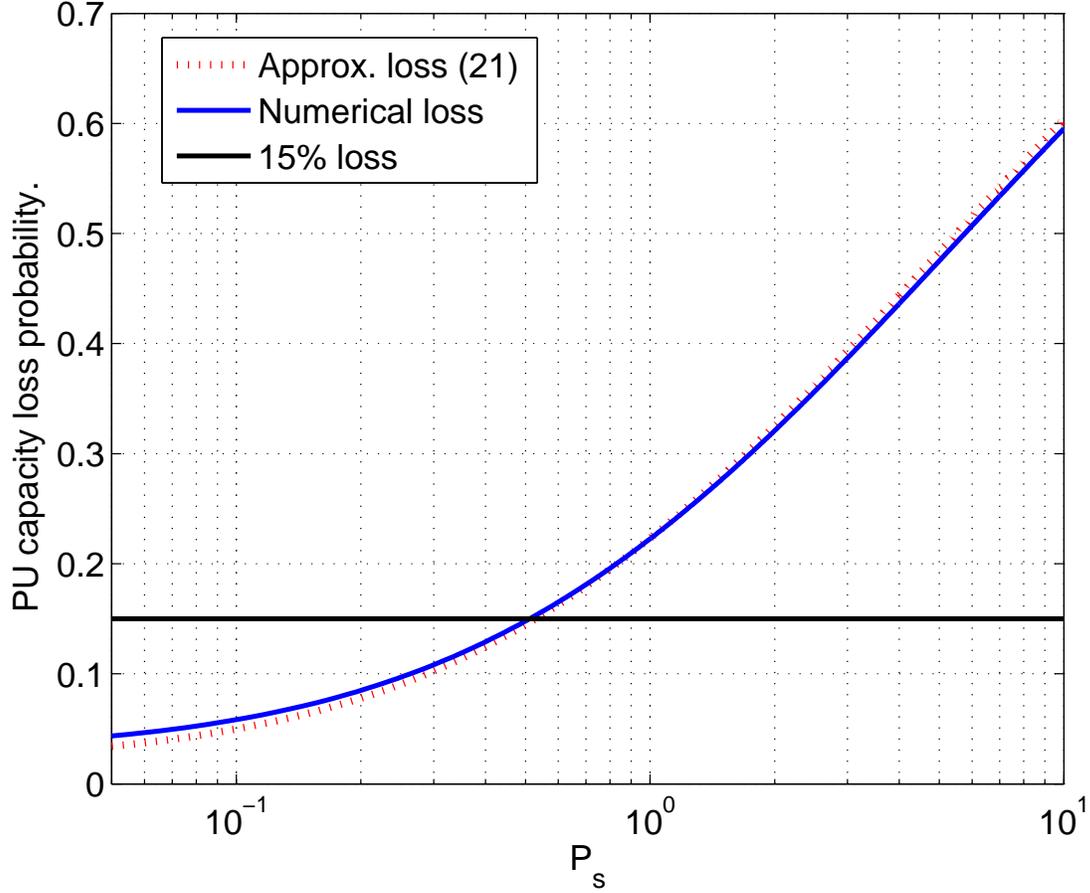}
	\caption{PU capacity loss (approx.) vs SU transmit power. $\gamma_p = 10$, $N_s=1$, $\overline{g}_0=1$.}
	\label{fig:pu_closs}
\end{figure}

Fig. \ref{fig:c_pu_apprx} illustrates (\ref{C_p_approx}), (\ref{C_p}), and the bounds in (\ref{C_p_bounds}). The approximated capacity loss of the PU can be obtained as
\begin{equation}
	C_{P,\text{approx}}^{\text{loss}} = C_P - C_P^{s,(\text{approx})} \label{Approx_los_prob}
\end{equation}
The approximated PU capacity loss probability is shown in Fig. \ref{fig:pu_closs}.

\section{PU capacity loss based power allocation} \label{sec:pu_c_los_pa}

\begin{figure}[htbp]
	\centering
		\includegraphics[width=0.90\columnwidth]{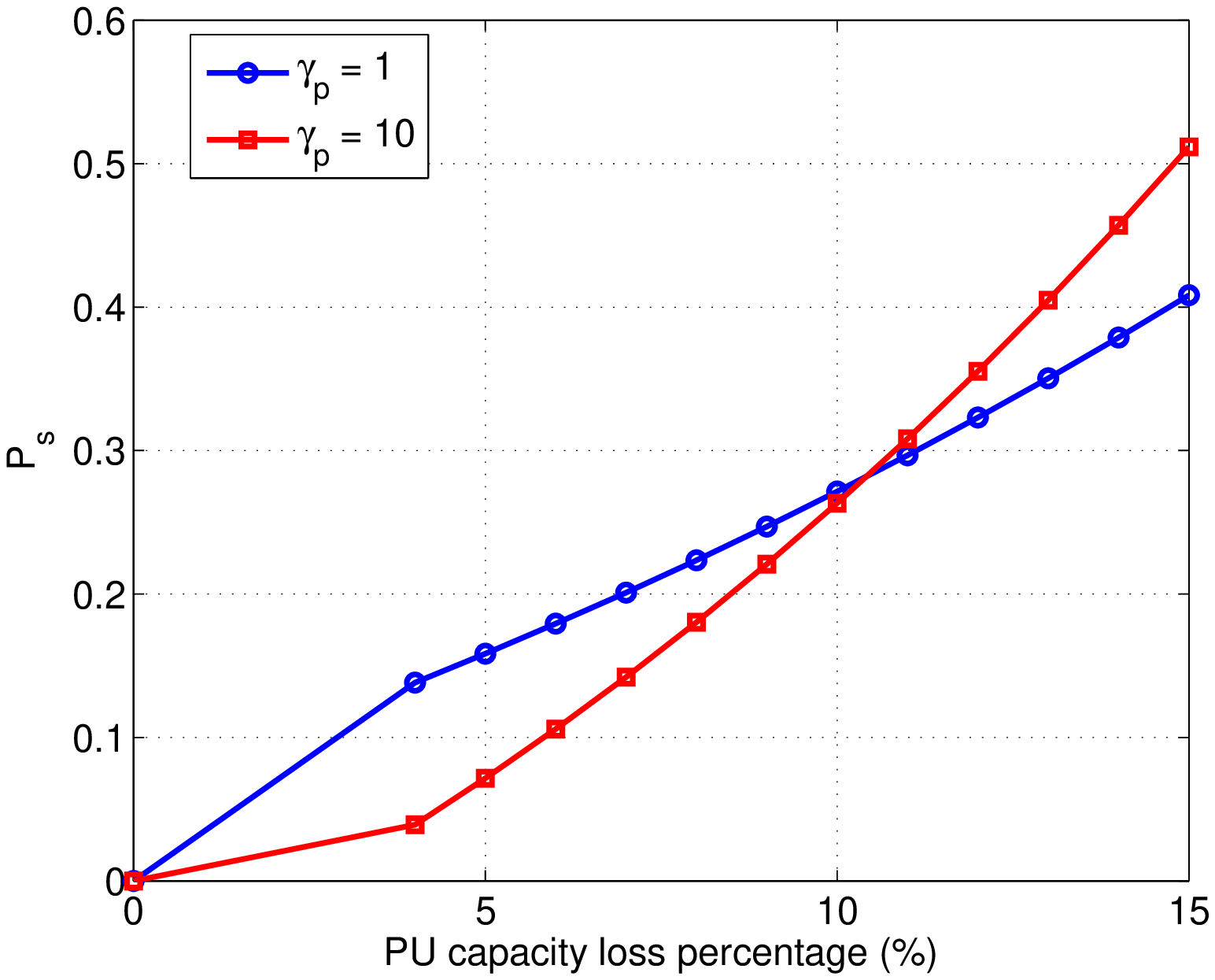}
	\caption{SU transmit power vs PU approximated capacity loss percentage (\ref{Approx_los_prob}). $N_s=1$ and $\overline{g}_0=1$}
	\label{fig:s_pow_closs}
\end{figure}

\begin{figure}[htbp]
	\centering
		\includegraphics[width=0.90\columnwidth]{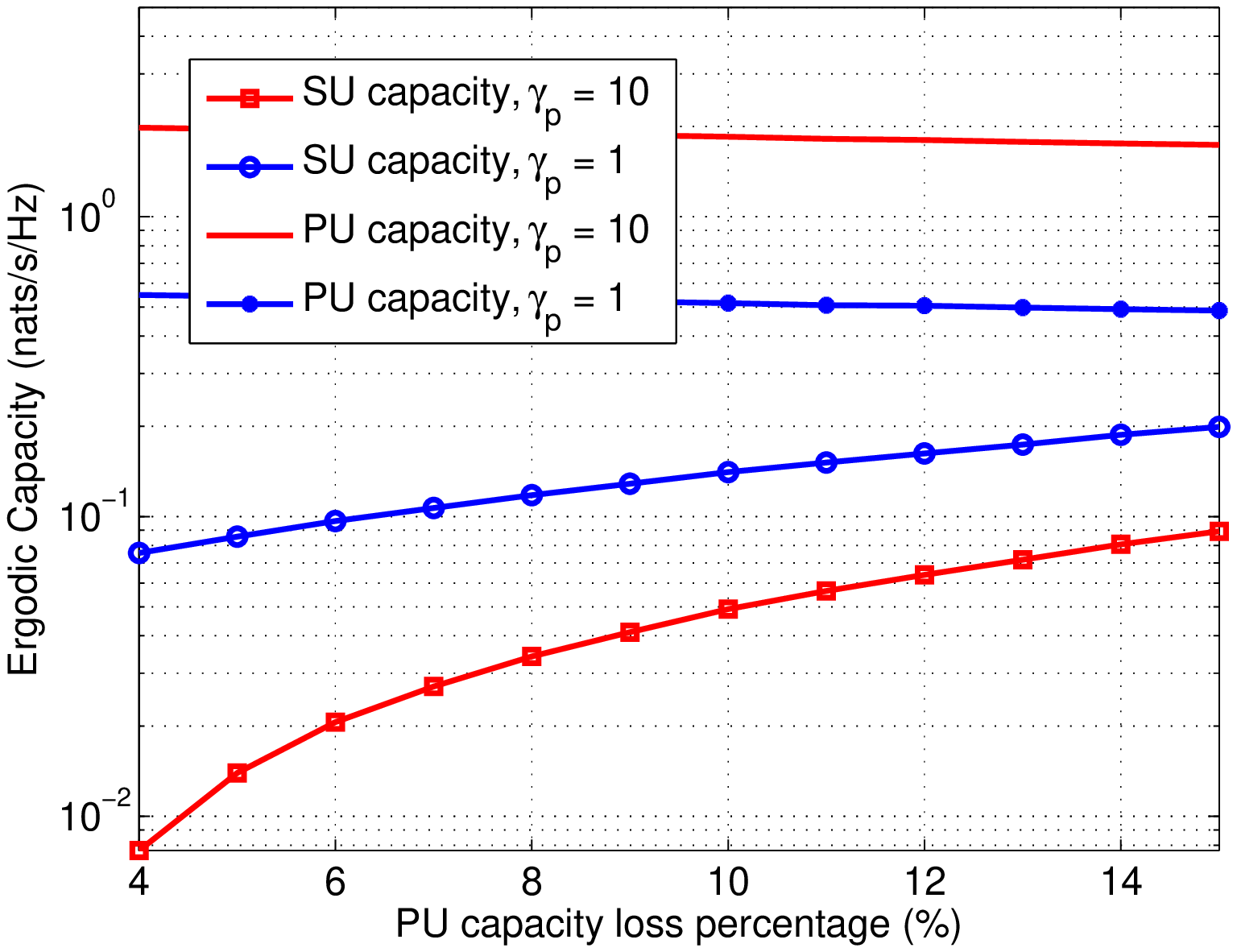}
	\caption{Ergodic capacity of the SU and PU, where the SU uses the proposed PU capacity loss based power allocation scheme.}
	\label{fig:c_sim_loss}
\end{figure}

In this section, we show the results of the PU capacity loss based power allocation scheme for the secondary user. It is obviously that the interference by the secondary transmission to the primary user always causes capacity loss to the PU. It is easy to numerically obtain the values for the secondary transmission power from (\ref{Approx_los_prob}) when the capacity loss probability is given. Fig. \ref{fig:s_pow_closs} shows the SU transmit power versus PU capacity loss percentage. Using these numerical results, we illustrate in Fig. \ref{fig:c_sim_loss} the ergodic capacity of the SU and PU, where the SU applies the proposed PU capacity loss based power allocation scheme.

There is an interesting finding observed from Fig. \ref{fig:s_pow_closs}, which has not been reported in the literature, that when the PU runs on the high SNR regime, the secondary transmission has much higher impact on the PU capacity loss percentage than that in the lower SNR regime, such that the SU applying based power allocation scheme needs to transmit with lower power. The comparison between Fig. \ref{fig:c_su} and Fig. \ref{fig:c_sim_loss} shows that the behavior of the PU transmission and the constraints lead to different scenarios for the SU ergodic capacity. When the PU works on higher SNR regime, PU capacity loss percentage based power allocation results in lower SU ergodic capacity than the case that the PU is in lower SNR regime. This is different from the interference outage probability constraint, where the interference threshold is defined as the percentage of the average value of the received desired signal strength at the PR, for instance, 10\%. In this scenario, higher SNR regime of the PU leads to higher SU ergodic capacity. 

\begin{figure}[htbp]
	\centering
		\includegraphics[width=0.90\columnwidth]{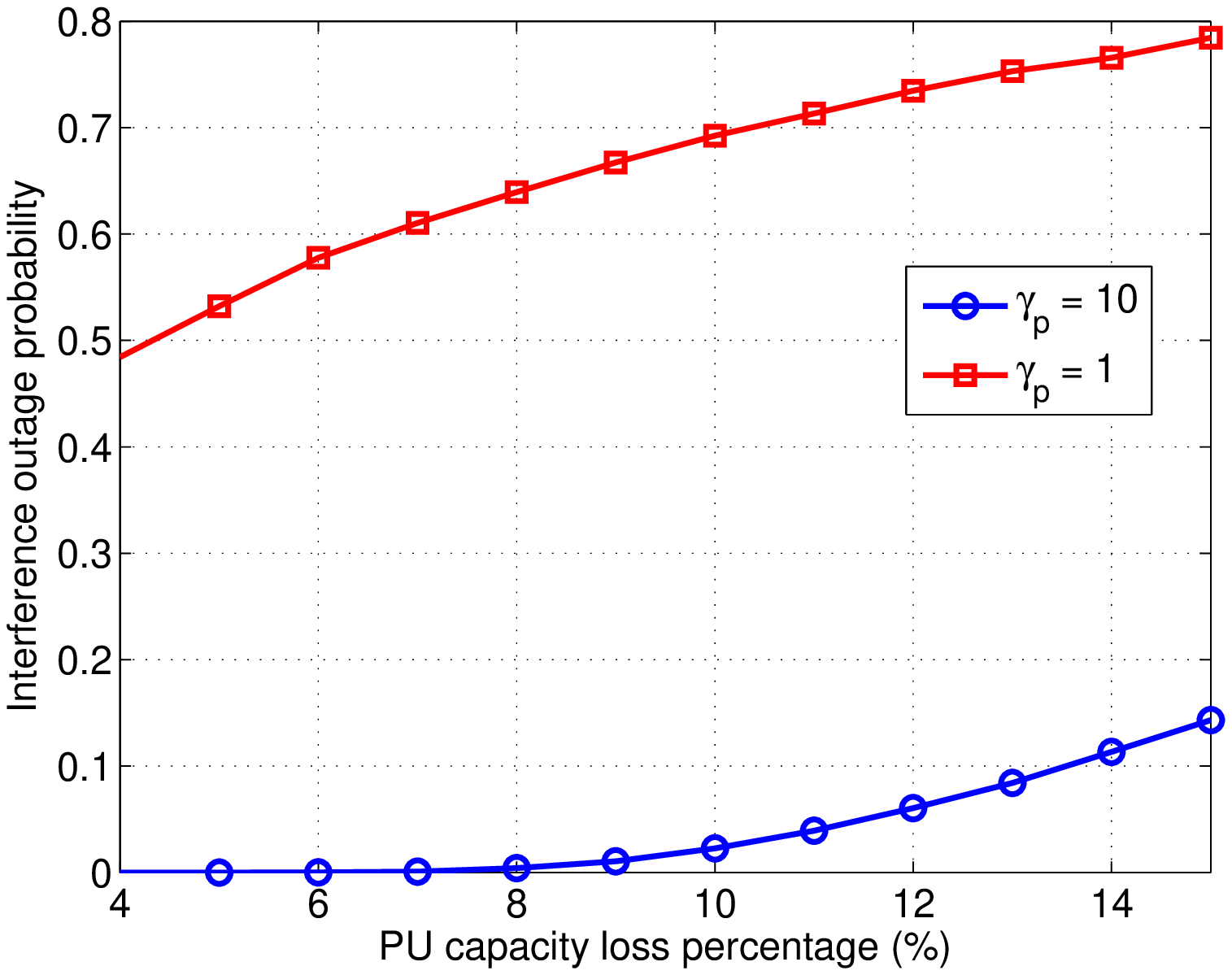}
	\caption{PU capacity loss percentage versus interference outage probability, where the interference threshold $I_{th} = 0.1 \gamma_p$.}
	\label{fig:loss_outage}
\end{figure}

Next, we depict the relation between the capacity loss percentage and the interference outage probability. Fig. \ref{fig:loss_outage} illustrates that the SNR regime of the PU significantly affects the SU performance. When the PU is working on higher SNR regime, the PU capacity loss leads to much lower interference outage probability at the PR, such that the SU achieves lower capacity than that when the PU is in the higher SNR regime. Although, the PU capacity loss percentage base power allocation scheme in lower SNR regime of the PU will result in higher capacity for the SU, it leads larger interference outage probability to the primary user. These results are importance of planning the coexisted systems when the perfect interference channel information is not available at the secondary transmitter.

\begin{figure}[htbp]
	\centering
		\includegraphics[width=0.90\columnwidth]{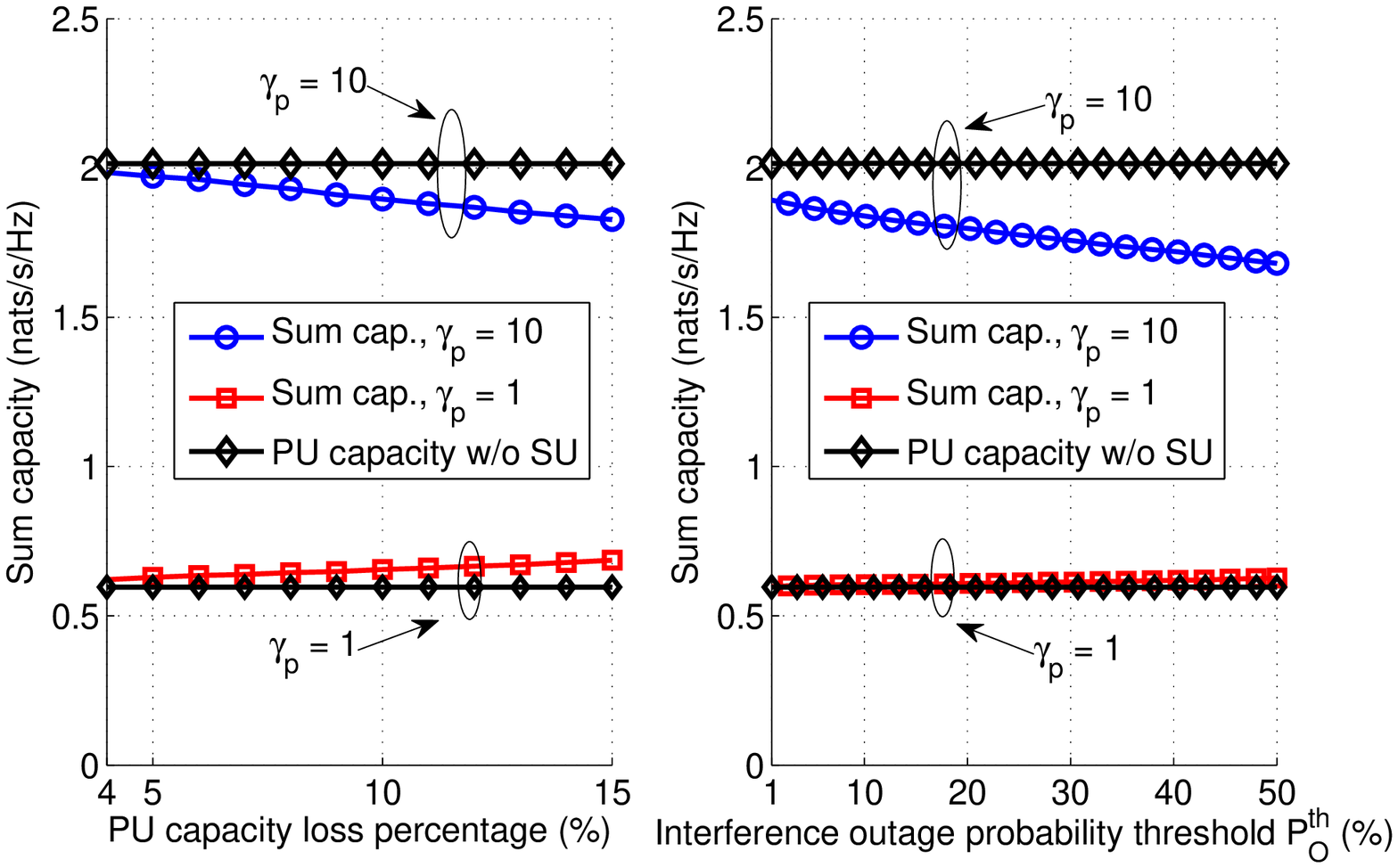}
	\caption{Sum capacity of the primary and the secondary users.}
	\label{fig:sum_cap}
\end{figure}

The sum capacities for the two power allocation schemes are shown in Fig. \ref{fig:sum_cap}. We have the following observations:
	
	\begin{enumerate}
		\item when the PU works at higher SNR regime, the sum capacity of the PU and SU is less than the value of the PU capacity (without secondary transmission);
		\item the sum capacity is decreasing while the constraints become looser;
		\item there is a difference phenomenon in lower SNR regime of the PU that the sum capacity increases in both schemes while the constraints becomes looser;
		\item in lower PU SNR regime, the increase of the sum capacity using PU capacity loss percentage base power allocation for the SU is slightly higher than the one using interference outage probability based power allocation; in higher PU SNR regime, the amount of sum capacity loss using PU capacity loss based power allocation scheme is also slightly less than the interference outage probability based power allocation.
	\end{enumerate}

\section{Conclusion and Discussion}

In this paper we have proposed an exact expression for the ergodic capacity of the secondary user with the consideration of the interference from the primary user, where the secondary transmission is constrained by the interference outage probability at the primary user. In order to have some explicit insights, a unified closed-form expression of the tight bounds for the secondary ergodic capacity has also been derived. For the purpose to understand the impact of secondary transmission on the ergodic capacity loss of the primary user, we have provided a pretty simple and accurate approximation and a closed-form expression for the primary user ergodic capacity loss. Through comparison between the interference outage probability based power allocation scheme and the primary capacity loss percentage based one, we have illustrated that for different secondary transmission strategies, the sum capacity are quite different.

The results of different power allocation associated to the different constraints (primary user requirements) can be used for system planning. For instance, the primary users located at the cell board usually have lower SNR than the PUs at the center of the cell. Therefore, there may have different constraints for SUs located at different area. The interference from the primary user to the secondary receiver may be applied by the secondary transmitter fed back from the secondary receiver to further improve the performance of the secondary user.

\ifCLASSOPTIONcaptionsoff
  \newpage
\fi

%
\bibliographystyle{IEEEtran}
\bibliography{E:/MyWork/MyResearch}

\end{document}